\documentstyle[12pt,epsfig]{article}

\newcommand{\beq}{\begin{equation} }
\newcommand{\eeq} {\end{equation} }

\begin{document}

\begin{flushright}
OSU-HEP-02-12\\
FERMILAB-PUB-02/308-T
\end{flushright}

\vspace*{2.5cm}
\begin{center}
{\Large {\bf Implications of a Massless Neutralino for Neutrino Physics \\
 } }

\vspace*{2cm}

I. Gogoladze\footnote{Email address:  ilia@hep.phy.okstate.edu}$^{,\dag,\ddag}$,
J. Lykken \footnote{Email address: lykken@fnal.gov }$^{,\ddag}$,
C. Macesanu\footnote{Email address:  mcos@pas.rochester.edu}$^{,\dag,\ddag}$,
and S. Nandi\footnote{Email address: shaown@okstate.edu}$^{,\dag,\ddag}$

\vspace*{0.5cm}
$^{\dag}${\it Department of Physics, Oklahoma State University\\
Stillwater, Oklahoma, 74078\\}

$^{\ddag}${\it Fermi National Accelerator Laboratory\\
P.O. Box 500, Batavia, Il 60510}\footnote{I.G., C.M. and S.N.
were participants in the Summer Visitor Program at Fermilab.}
\end{center}

\vspace*{0.5cm}

\begin{abstract}
We consider the phenomenological implications of a soft SUSY
breaking term $\tilde{B} N$ at the TeV scale (here $\tilde{B}$ is
the $U(1)_Y$ gaugino and $N$ is the right-handed neutrino field).
In models with a massless (or nearly massless) neutralino, such a
term will give rise through the see-saw mechanism to new
contributions to the mass matrix of the light neutrinos.
 We treat the massless neutralino as an (almost) sterile neutrino and find that its mass depends  on the square of
the soft SUSY breaking scale, with interesting consequences for
neutrino physics. We also show that, although it requires
fine-tuning, a massless neutralino in the MSSM or NMSSM is not
experimentally excluded. The implications of this scenario for
neutrino physics are discussed.
\end{abstract}

\newpage
\section{Introduction}

The atmospheric neutrino data \cite{ref1} gives convincing
evidence of non-zero neutrino masses. This data also implies maximal
or close to maximal mixing of muon and tau neutrino.
 Furthermore, the solar neutrino data \cite{ref2} can also be naturally explained by
non-zero mass splitting and mixing between the electron and muon
neutrino.
 Recent data favors large mixing in this sector as well.

The question arises then, what is the physics behind neutrino masses and
mixing patterns?
What is the mechanism of
neutrino mass generation? Why is the lepton mixing large,
in contrast to the small mixing in the quark sector?
These are among the most
challenging problems of fundamental physics today.

Many mechanisms for neutrino mass generation have been suggested
so far. Among these, the see-saw mechanism \cite{sees} seems to be
the simplest and most natural. In this context, large neutrino
mixing can appear due to large mixing in the charged lepton mass
matrix~\cite{ref5} or in the Dirac mass matrix of
neutrinos~\cite{ref6}. It can also appear from large mixing or
very strong hierarchy in the Majorana mass matrix of the right
handed neutrinos~\cite{ref7}. Large  mixing in the lepton sector
can also be obtained by radiative corrections due to
renormalization group effects in schemes with quasi-degenerate
neutrinos~\cite{ref8}.

Models with three light neutrinos can explain the solar and atmospheric
neutrino oscillation data, and also
all laboratory neutrino experiments results, with the exception of
 LSND \cite{LSND}.
 The explanation of the LSND experimental results
requires either a fourth light neutrino ~\cite{sss1,balaji} (which, to
satisfy the constraints coming from Z physics, has to be sterile),
or violation of CPT in the neutrino sector ~\cite{cpt}. In this
later case, the neutrino and antineutrino masses can be different,
thus providing an elegant solution to the LSND question. However,
violation of CPT may be hard to accommodate theoretically. In the
sterile neutrino case, one of the problems is that there is no
 compelling reason for the existence of such a particle. Also,
even if the sterile neutrino is introduced by hand, it is hard to
find a reason why it is so light, with mass of the order 1 eV, as
required to explain LSND.

In this work we propose a scenario in which the sterile
 neutrino is an essentially massless (mass
of order eV or less) bino. It can be shown that, in the framework
of the general Minimal Supersymmetric Standard Model (MSSM),
such a massless neutralino
is still allowed experimentally. In order to couple this
neutralino to the neutrino sector, we consider a new
soft SUSY breaking term of the form $c \tilde{B} N$, where $N$ is the
right handed neutrino and $c$ is a soft
SUSY breaking mass parameter.
 This coupling will give the bino mass through the see-saw mechanism:
$m_{\nu_0} \sim c^2 / M_M$, thereby relating  the mass
of the sterile neutrino to the soft SUSY breaking scale.
Moreover, the see-saw induced couplings of this bino with the
three SM neutrinos naturally leads to large mixing between the three active
families.

The paper is organized as follows. In sect. II we show how a
massless neutralino can become the sterile neutrino.
We also discuss the connection between the neutrino mass and the
SUSY breaking scale, and how large mixing arises naturally in this model.
  In sect. III we formulate
and review the conditions under which the lightest neutralino can
be massless (both in MSSM and in NMSSM), and satisfy all the
existing experimental constraints.  A new $U(1)_R$ symmetry is
introduced in order to allow the bino-neutrino (or
singlino-neutrino, in NMSSM) couplings, while forbidding all the
usual R-parity violating (RPV) terms.
 In sect. IV we consider the implications of our model for
neutrino phenomenology in somewhat more detail. Conclusions are
summarized in sect. V.

\section{See-saw mechanism and the massless neutralino}

In this section we will assume the existence of a massless
neutralino, leaving the justification for this assumption for
later. This neutralino can be thought of as a superposition of
higgsino and gaugino states. In the next section we will see that
in the MSSM  it has to be mostly bino in order to satisfy the
existing experimental constraints.

Let us consider scenarios under which mixing between this state and
the three SM neutrinos can arise.
 The first possibility is direct mixing: the
higgsino states can couple to the neutrinos through bilinear RPV
couplings $\mu_i$; we will not study this scenario here. We shall
consider the more interesting case when the mixing with the three
SM neutrinos is obtained through coupling of the massless
neutralino to the right-handed neutrino and the see-saw mechanism.

 We take the right handed neutrino fields $N_i$ to be singlets
under the $SU(2)_{L}\times U(1)_Y$ gauge group. Then they can
couple only with the bino component of the massless neutralino,
and to $SU(2)_L \times U(1)_Y$ singlet components of the
neutralino, if present. The corresponding bilinear couplings to
the bino:
\begin{equation}\label{bb}
 \tilde{B} N_i
\end{equation}
have dimension three, and are therefore soft SUSY breaking terms
~\cite{grish}. The
 coefficients with which these terms appears in the Lagrangian
(let's call them $c_i$) will have dimensions of mass, and their magnitude is
expected to be of the order of SUSY breaking scale, that is, ${\cal O}$(TeV).

The Lagrangian for the neutrino sector of our model will then be:
\begin{equation}\label{lag}
   L=(m_D)_{ij} \bar{\nu}_i  N_j + (M_M)_{i j} N_i  N_j +
   c_{i} N_i \chi_0 + h.c.,
\end{equation}
where $m_D$ and $M_M$ are the Dirac and Majorana mass matrices for
the three SM neutrinos, and $\chi_0$ is the massless neutralino.
We treat the massless neutralino as an (almost) sterile neutrino.
Upon decoupling the heavy neutrino states (through the see-saw
approximation) the mass matrix for the remaining four light
neutrinos takes the form:
\begin{equation}\label{nu_mm}
M_\nu = \left(\begin{array}{cc}
         c M_M^{-1} c^T & c M_M^{-1} m_D^T \\
         m_D M_M^{-1} c^T & m_D M_M^{-1} m_D^T
       \end{array}\right)
\end{equation}
where the first line corresponds to the sterile neutrino.

For simplicity  of presentation let's assume in the following that
the right handed neutrino
mass matrix is diagonal and proportional to the identity matrix:
$(M_M)_{ij}= m_M \delta_{ij} $. It can easily be seen then that the
 4-neutrino mass matrix above
has a zero eigenvalue, two eigenvalues of order $m_D^2/m_M$ and
one eigenvalue of order $(c^2 + m_D^2)/m_M$. The first three
eigenvectors can be identified with the three SM neutrino mass eigenstates.
The fourth eigenvector (which is mostly $\chi_0$) can be identified
with the sterile neutrino.

Note that, since the magnitude of the $c$ terms is of order of
soft SUSY breaking scale, they are naturally about ten times
larger than the Dirac mass terms appearing in $m_D$ (which are  of
the order of the electroweak breaking scale, {\it i.e.} ${\cal
O}$(100 GeV)). That means that the mass of the sterile neutrino
will be about two orders of magnitude above the mass of the
heavier SM neutrino, which, to account for the atmospheric
neutrino data, has to be of order $5\times 10^{-2} $eV. This makes
the mass of the sterile neutrino of eV size, which is the right
value to explain the LSND experiment results. Moreover, we will
show in  section IV that Eq. (\ref{nu_mm})
 predicts the right value of mixing
between the sterile and active neutrinos as well.

Notwithstanding the LSND experiment, the mass matrix in Eq.
(\ref{nu_mm}) with $|c| \gg |m_D|$  naturally gives
large mixing between the three light neutrinos.
To see this we can decouple the sterile neutrino from
the other three by using the see-saw approximation, and the mass matrix becomes:
\begin{equation}\label{nu_mm1}
 M_\nu' \ = \ \frac{1}{m_M}
\left(\begin{array}{cc}
         c  c^T & 0 \\
         0  & (m_D m_D^T)_{ij} -
         { \displaystyle \frac{(c m_D^T)_i \ (c m_D^T)_j}{c c^T} }
       \end{array}\right).
\end{equation}
From this expression it can be seen that even if $m_D$ is
diagonal, we get off-diagonal entries  of the same magnitude as
the diagonal elements in the three SM neutrino mass matrix. This
means that at least  one mixing angle is large. Of course, in
order to obtain the specific pattern of two large mixing angles
and a small one, further conditions must be imposed. We will study
this further in chapter IV.

We end this section with some comments. Above, we have assumed
that the sterile neutrino is a massless neutralino. This does not
necessarily have to be so. What is needed for the above see-saw
scenario to work is a massless fermion, which is (or contains
among its components) a singlet under the $SU(2)_L\times U(1)_Y$
gauge group. Then, this fermion can couple with the right handed
neutrinos through the soft SUSY breaking terms (\ref{bb}), and the
mechanism presented above works. In the next section we will
actually consider the case when the sterile neutrino is the NMSSM
singlino. Some other SUSY particles (like a goldstino, which has
the advantage of being naturally massless) can also play this
role.

\section{Massless Neutralino}

In this section we will explore the possibility that supersymmetry
allows the existence of a nearly massless neutralino. For the MSSM
and the NMSSM, we show that a massless neutralino can be obtained
by a fine-tuning of the soft breaking parameters. While we do not
provide a reason for such a tuning, we do verify that the
resulting massless neutralino is not yet excluded by experiment.
By extending the visible sector particle content beyond that of
the MSSM or NMSSM, it may be possible to achieve TeV scale visible
sector SUSY breaking in a phenomenologically viable way. In such a
case the lightest neutralino can be the goldstino and thus
naturally massless, up to supergravity corrections of order
TeV$^2/M_{\rm Planck}\sim 10^{-4}$ eV. This is an interesting
direction for further study.

In this section we will explore the possibility that supersymmetry allows
the existence of a massless neutralino. Note that we will not try to
give a reason why there should be such a particle
(at this point, we do not know), but we will just simply verify that
the existence of a massless neutralino is compatible with
the experimental results from SUSY searches. We shall consider two
SUSY models, first the MSSM, and then the next-to-minimal
supersymmetric model (NMSSM), which contains an extra singlet.

\subsection{MSSM case}

 In the MSSM we have the following mass matrix for neutralinos:

\begin{equation}
\label{n1} M_{ij}\equiv\pmatrix{-M_1 &0 &-m_z\cos\beta\sin\theta_w
&m_z \sin \beta \sin \theta_w \cr \nonumber 0 & -M_2  &m_z \cos
\beta\cos\theta_w & -m_z \sin \beta \cos\theta_w\cr\nonumber
 - m_z\cos\beta\sin\theta_w & m_z \cos\beta\cos\theta_w & 0 &-\mu\cr\nonumber
  m_z \sin \beta \sin \theta_w &- m_z \sin \beta \cos \theta_w & -\mu & 0}
\end{equation}
where $M_1$ and $M_2$ are the soft SUSY breaking mass terms for the
 $U(1)_Y$ and $SU(2)_w$  gaugino fields, $m_z$ is the mass of the
Z-boson, $\tan\beta\equiv v_1/v_2$ and $\mu$ is the Higgs mixing term.

 The existence of a massless neutralino requires that
the determinant of the mass matrix in (\ref{n1}) be zero:
\begin{equation}\label{n4}
\Delta_0  \equiv \mu m_z^2 \sin2\beta (M_1\cos^2\theta_w+ M_2
\sin^2\theta_w) -\mu^2 M_1 M_2 \ = \ 0
\end{equation}
 Most studies of the MSSM have been performed with the
assumption of universal gaugino masses, where $M_1$ and $M_2$ are
related by the GUT relation:
\begin{equation}\label{GUT}
 M_2 = \frac{5}{3} \tan^2 \theta_W M_1
\end{equation}
In this framework, Eq. (\ref{n4}) requires that:
\begin{equation}\label{d02}
 \mu M_2 =\frac{ m_z^2}{r} \sin2\beta (r \cos^2\theta_w + \sin^2\theta_w)
\end{equation}
with $r=M_1/M_2 \simeq 0.5$, which implies either that both $\mu$
and $M_2$ are of order $m_z$, or one of these parameters are much
smaller than $m_z$. However, since $\mu$ and $m_2$ are responsible
for the masses of the other neutralinos, as well as for the masses
of the charginos, this can bring us into conflict with direct
searches for these particles at LEP. A detailed analysis
~\cite{ster} shows that a massless neutralino is excluded in the
MSSM with the GUT relation (\ref{GUT}), except for a narrow region
in the parameter space, where $\tan \beta \simeq 1$. However, this
value for $\tan \beta$ is excluded from other considerations.

  In the search for a massless neutralino we are  therefore led to
give up the assumption of universal gaugino masses. Then, values
for the $M_1$ parameter:
\begin{equation}\label{m1exp}
M_1 = \frac{M_2 m_z^2 \ \sin2\beta \sin^2\theta_w} {\mu M_2 - 
m_z^2 \sin2\beta \cos^2\theta_w} \simeq \frac{m_z^2}{\mu}
\sin2\beta \sin^2\theta_w
\end{equation}
of order of a few GeV (or even smaller, for large $\tan \beta$),
can satisfy Eq. (\ref{n4}). $M_2$ and $\mu$ can be chosen
sufficiently large to satisfy the constraints coming from $Z$
decay to $\chi^+ \chi^-$ and $\chi^0_i \chi^0_j$, with $i$ and $j$ not 1 at
the same time.
One more constraint we have to consider comes from the massless
neutralino, which  will give contribution to the invisible Z
width. The current experimental \cite{pdg} value:
\begin{equation}\label{n3}
    \Gamma_Z^{inv} = 499.0 \pm 1.5~ \hbox{MeV}
\end{equation}
requires that the branching ratio of $Z$ to the massless
neutralino pair be smaller than about  0.3 \%:
\begin{equation}\label{z_inv}
\hbox{Br}.(Z \rightarrow \chi^0_1 \chi^0_1) < 3 \times 10^{-3} .
\end{equation}

In order to figure out this branching ratio, we need to evaluate
the particle content of the lightest neutralino and its
interactions. If we define the mass eigenstates of the neutralino
matrix by:
\begin{equation}\label{rot}
\chi^0_i = N_{ij} \psi_j \ \ \  , \ \  \hbox{where } \psi_j = \{
\tilde{B},\tilde{W^3},\tilde{h^0},\tilde{h'^0} \},
\end{equation}
then
\begin{equation}\label{nu_con}
N_{1i} = (\ 1\ ,\ \frac{m_z^2 \sin2\beta \sin2\theta_W /2}{m_z^2
\sin2\beta \cos^2\theta_W -\mu M_2}\ , - \frac{m_z M_2 \sin\beta
\sin \theta_W}{m_z^2 \sin2\beta \cos^2\theta_W -\mu M_2} \ ,
\frac{m_z M_2 \cos \beta \sin \theta_W}{m_z^2 \sin2\beta
\cos^2\theta_W -\mu M_2} \ )
\end{equation}
up to a normalization constant. If $\mu \gg m_z$, that
normalization constant is 1, and the massless neutralino is mostly
bino. The interaction Lagrangian  with the Z boson in terms of
physical states is the following \cite{hk}:
\begin{equation}\label{n5}
    {\cal L}_{Z\chi_i\chi_j}=
    (\frac{g}{2\cos\theta_w}) Z_{\mu}\bar{\chi_i}\gamma^{\mu}
    (O_{ij}^{L} P_L+ O_{ij}^{R},
    P_R)\chi_j
\end{equation}
where
\begin{equation}\label{n6}
O_{ij}^{L}=\frac{1}{2}(- N_{i3}N_{j3}^\ast+N_{i4}N_{j4}^\ast),
\hspace{1.cm} O_{ij}^{R}=-O_{ij}^{L \ast} .
\end{equation}
Therefore, for the massless neutralino we have:
 \begin{equation}\label{n7}
O_{11}^{L} \simeq \frac{1}{2} \frac{m_Z^2}{\mu^2} \sin^2\theta_W
\cos2\beta .
\end{equation}
Eq. (\ref{z_inv}) requires that $O_{11}^L \le 1/30$; we see that
this can be easily satisfied for values of $\mu$ of order 500 GeV.
(Note also that although in the limit $\tan\beta \approx 1$ the
massless neutralino is decoupled from Z-boson this is not a
necessary condition).

So far we have shown that a massless neutralino is consistent with
experimental constraints (if we give up the assumption of gaugino
mass universality at GUT scale). This massless state, however,
does not appear naturally in the theory; fine tuning of order 1
eV/ 100 GeV $\simeq 10^{-11}$ is necessary to satisfy Eq.
(\ref{n4}). Moreover, we have to impose this fine tuning on the
complete theory; for example, if Eq. (\ref{n4}) holds at tree
level, it will be broken by loop corrections, and the neutralino
will acquire GeV size mass. For these reasons, this model is not
very compelling theoretically; however, it is experimentally
allowed.

\subsection{NMSSM case}

In the event we want to keep the GUT relation (\ref{GUT}), it is
possible to get a massless neutralino in the framework of NMSSM.
Let us  considering NMSSM with the following terms in the superpotential:
\begin{equation}\label{n8}
    W = \lambda\varepsilon_{ij}H_1^i H_2^j S -\frac{1}{3}\kappa S^3
\end{equation}
where $i,j=1,2$, $\epsilon_{ij}$ is the antisymmetric tensor, $H_1$
and $H_2$ are the standard Higgs doublets and $S$ is the MSSM  Higgs
singlet. (The first term replaces the usual SUSY $\mu$ term).
 In this case the neutralino mass matrix becomes:

\begin{eqnarray}
\label{n11} M_{ij}\equiv\pmatrix{-M_1 &0
&-m_z\cos\beta\sin\theta_w &m_z \sin \beta \sin \theta_w & 0\cr
\nonumber 0 & -M_2  &m_z \cos \beta\cos\theta_w & -m_z \sin \beta
\cos\theta_w & 0 \cr\nonumber
 - m_z\cos\beta\sin\theta_w & m_z \cos\beta\cos\theta_w &
  0 &\lambda x & \lambda v
 \sin\beta\cr\nonumber
  m_z \sin \beta \sin \theta_w &- m_z \sin \beta
 \cos \theta_w a &  \lambda x & 0 & \lambda v \cos\beta
\cr\nonumber
  0 & 0  & \lambda v \sin\beta & \lambda v \cos\beta &
-2\kappa x }
\end{eqnarray}
The determinant of this mass matrix is:
\begin{equation}\label{m1}
\Delta=-2 \kappa x \Delta_0+\lambda^2v^2(m_z^2(M_1\cos^2\theta_w +
    M_2\sin^2\beta_w) - \mu M_1M_2\sin2\beta)
\end{equation}
were $v$=174 GeV, $x$ is vacuum expectation value (VEV) of $S$
field, and we define $\mu$ to be $\lambda x$. We assume in
the following that we are in a region of the parameter space where
$\mu M_1 M_2 \gg m_z^2 (M_1 \cos^2\theta_W +M_2 \sin^2\theta_W)
\simeq 0.6 m_z^2 M_2$; we also assume that $\tan \beta$ is
large. Then $\Delta_0 \simeq -\mu^2 M_1 M_2$, and:
\begin{equation}\label{k_val}
\kappa = \lambda \frac{1}{2} \left( \frac{\lambda v}{\mu}
\right)^2 \frac{0.6 m_z^2 M_2 - 0.5 \mu M_2^2 \sin2\beta}{- \mu
M_1 M_2}
\end{equation}
will provide a massless neutralino. The particle content of this
neutralino is given by:
\begin{equation}\label{mas_nu}
N_{1i} = \left(\ \lambda \frac{ v\ m_z }{\mu M_1} \cos2\beta
\sin\theta_W , -\lambda \frac{ v\ m_z }{\mu M_2} \cos2\beta
\cos\theta_W ,\right.
\end{equation}
$$
\left. -\lambda \frac{v}{\mu} \frac{0.6  m_z^2 \sin\beta - \mu M_1
\cos\beta} { \mu M_1} , -\lambda \frac{v}{\mu} \frac{0.6 m_z^2
\cos\beta - \mu M_1 \sin\beta} { \mu M_1} ,1
 \right)
$$
that is, in the limit when $\lambda v / \mu \ll 1$, it is mostly
singlino. The coupling to the Z is given by:
\begin{equation}\label{Z_cp}
O^L_{11} \simeq \left(  \frac{\lambda v}{\mu} \right)^2 \
\frac{(\mu M_1)^2-(0.6  m_z^2)^2}{(\mu M_1)^2}\simeq \left(
\frac{\lambda v}{\mu} \right)^2
\end{equation}
and with $\mu$ of order 500 GeV and $\lambda \simeq 0.3$, the
constraint coming from the invisible Z width is satisfied again.
(Note that this implies that $\kappa$ is quite small as well).

\subsection{The $c$ terms}

We have seen that  bi-linear terms (\ref{bb}) which couple the
bino (or, in the NMSSM case, the singlino) and the right handed
neutrino are allowed by the requirement that supersymmetry is
softly broken. However,  these terms break usual $R$ parity, which
may not be desirable. In order to forbid the usual RPV terms, we
can introduce a new $U(1)_R$  symmetry.
 Let us then consider the following
generation independent assignment of $U(1)_R$ charges to
the MSSM and right handed neutrino superfields:
\begin{equation}
\label{mssm}
\begin{array}{cccccc}
  Q &(3,2,1/6) : & q~~~ & ~~~D^c&({\bar 3}, 1,1/3): & 2-2q-u \\
  U^c&(\bar{3},1,-2/3): & u~~~ & ~~~L& (1,2,-1/2): & -2+q+u \\
   N&(1,1,0) : & 2~~~ & ~~~E^c& (1,1,1) : & 4-2q-2u \\
     H_d&(1,2,-1/2) : &  q+u~~~ & ~~~H_u&(1,2,1/2) : & 2-q-u  \\
   \theta&  : & 1 & &
\end{array}
\end{equation}
 where   the
$SU(3)_c\times SU(2)_L\times U(1)_Y$ quantum numbers of the
particles are  given in the brackets,
and $\theta$ is the Grassmann coordinate. We choose the
$U(1)_R$ charge of the Grassmann coordinate $\theta$ to be unity.
 It is easy to check  that this charge assignment forbids
the usual  R-parity breaking terms and allows all
MSSM Yukawa couplings as well as the right handed neutrino
$\widetilde{B}N$ coupling and the Higgs mixing  $\mu$ term. In our
$U(1)_R$ charge approach we assume that the right handed neutrino
fields have $U(1)_R$ charge equal to $-1$; this means that the
corresponding mass terms are
generated though a heavy field VEV ($
\ge 10^{14}$ GeV). This field  also breaks the $U(1)_R$ symmetry at the high
scale.

In the NMSSM case, when the massless neutralino is mostly
singlino, the relevant coupling  with the right handed
neutrino is:
\begin{equation}\label{sing}
  c_i \widetilde{S} N_i
\end{equation}
This  coupling also breaks usual $R$ parity. But we can choose the
following $U(1)_R$ charge assignments:
\begin{equation}
\label{nmssm2}
\begin{array}{cccccc}
  Q &(3,2,1/6) : & q~~~ & ~~~D^c&({\bar 3}, 1,1/3): & \frac{8}{3}-2q-u \\
  U^c&(\bar{3},1,-2/3): & u~~~ & ~~~L& (1,2,-1/2): & -\frac{10}{3}+q+u \\
   N&(1,1,0) : & \frac{10}{3}~~~ & ~~~E^c& (1,1,1) : & 6-2q-2u \\
     H_d&(1,2,-1/2) : &  -\frac{2}{3}+q+u~~~ & ~~~H_u&(1,2,1/2) : & 2-q-u  \\
   \theta&  : & 1 & S& \frac{2}{3}
\end{array}
\end{equation}
and in this case only the (\ref{sing}) couplings  are allowed from
general  the RPV term.

\section{Implications for neutrino sector}

In this section we will consider the implications of a  sterile
neutrino coming from supersymmetry for neutrino physics.

\subsection {LSND result and massless neutralino }

As was mentioned briefly in sect. II, a massless neutralino seems
ideally suited to explain the LSND experiment results.  Since
the magnitude of $c$ couplings is given by the SUSY breaking scale,
 we can expect
the mass of the fourth neutrino to be
of order $c^2/m_D^2 \simeq 100$ times larger than the mass of the
heaviest SM neutrino.
This puts it in the eV range, which is the right value
needed to account for LSND result.

 Moreover, LSND data \cite{LSND}
and  constraints from short baseline experiments
\cite{bugey,CHDS} requires that the admixture
of $\nu_e$ and $\nu_{\mu}$ in the fourth neutrino (let's call it $\nu_0$) has
to be of order of 0.1. In terms of the elements of the rotation matrix
$U_{\alpha i}$ ($\alpha$ stands for $\chi^0_1, \nu_e , \nu_{\mu}, \nu_{\tau}$)
which diagonalizes the $(M_{\nu})_{\alpha \beta}$ 4x4 mass matrix (\ref{nu_mm}),
we need $U_{e 0}, U_{\mu 0} \sim 0.12 - 0.14$. On the other
hand, as long as $M_{\nu}(\chi^0_1,\nu_i) \ll M_{\nu}(\chi^0_1,\chi^0_1)$
we can employ the see-saw approximation to decouple the fourth neutrino from
the other three, and we will have:
$$
 U_{i0} \simeq \frac{M_{\nu}(\chi^0_1,\nu_i)}{M_{\nu}(\chi^0_1,\chi^0_1)}
 \simeq \frac{m_D}{c} \simeq \frac{1}{10}
$$
which fits nicely the experimental requirements.

For purposes of illustration, we present here a particular
realization of this situation. We can work in a basis where the
Majorana mass matrix is diagonal; let's even assume that it is
proportional to the identity matrix: $M_M = I_{3\times 3} m_M$. Moreover,
let's assume that there is no hierarchy between the soft SUSY
breaking parameters: $c_1 = c_2 = c_3 = c$. For the Dirac mass matrix, take
 a symmetric form:
\begin{equation}\label{gm_d}
m_D = \left(
\begin{array}{ccc}
m_1 & a_3 & a_2 \\
a_3 & m_2 & a_1 \\
a_2 & a_1 & m_3
 \end{array}
\right)
\end{equation}
and see what constraints the LSND result imposes on the elements of this matrix.

First, we need $U_{e 0} \simeq U_{\mu 0}$; since in the see-saw
approximation $U_{i0} = M_{\nu}(\chi^0_1,\nu_i) / M_{\nu}(\chi^0_1,\chi^0_1)$,
  this implies $ \sum_i (m_D)_{i1} = \sum_i (m_D)_{i2}$, or:
\begin{equation}\label{aeq_1} m_1 + a_2 + a_3 = m_2 + a_1 + a_3 .
\end{equation}
Moreover, from atmospheric oscillations results and the CHOOZ
constraints on $\bar{\nu_e}$ disappearance \cite{CHOOZ}
($\theta_{23} \simeq 45^o, \theta_{13}$ small), we know that
$|U_{\mu 3}| \simeq |U_{\tau 3}| \simeq 1/\sqrt{2}$ and $|U_{e 3}|
\simeq 0$, which implies $|U_{0 3}| \simeq 0$, too. From the
orthogonality of the rotation matrix $ \sum_i U_{i 0} U_{i 3} =
0$, we then get $ |U_{\mu 0}| \simeq |U_{\tau 0}|$, or
\begin{equation}\label{aeq_2} m_2 + a_1 + a_3 = m_3 + a_1 + a_2 .
\end{equation}
After decoupling the sterile neutrino, the effective mass matrix
for the three SM neutrinos will be:
\begin{equation}\label{mnp}
(M_{\nu}')_{i j} = \frac{ (m_D m_D^T)_{ij} }{m_M} -
\frac{(M_{\nu})_{i0} (M_{\nu})_{0j}}{(M_{\nu})_{00}} .
\end{equation}
Since the determinant of this matrix is zero, there are three
 possible textures which will explain
the observed neutrino mass splitting and mixings (see, for example,
\cite{Babu_bm}).
 We shall try to obtain
the hierarchical form, where $(m_{\nu_1},m_{\nu_2},m_{\nu_3}) =
(0,\delta, M)$, with $\delta \ll M$.
 Then the mass matrix should look like (\ref{hier}).
Requiring that $ |(M_{\nu}')_{12}| \simeq |(M_{\nu}')_{13}| \simeq 0$
we get the following equations:
\begin{equation}\label{m_eqs}
m_2 = m_3 \ \ , \  a_1 = 2 m_1 -m_3.
\end{equation}
Then:
\begin{equation}\label{mm_o0}
 M_{\nu}' = \frac{1}{m_M}
\left(
\begin{array}{ccc}
0 & 0 & 0 \\
0 & 2(m_1-m_3)^2 & -2(m_1-m_3)^2 \\
0 & -2(m_1-m_3)^2 & 2(m_1-m_3)^2
 \end{array}
\right)
\end{equation}
leads to maximal mixing for the atmospheric neutrinos, with
the corresponding mass scale given by $M = 4(m_1-m_3)^2/m_M$. The
$\theta_{13}$ angle is also zero. However this texture does not explain
the solar oscillations. To account for these, we need corrections of order
$\delta/M$ to the texture (\ref{mm_o0}), where $\delta$ is the mass scale
responsible for solar oscillations. We are therefore led to consider
 corrections of order $\delta/M$ to Eqs. (\ref{aeq_1}, \ref{aeq_2}, \ref{m_eqs}).
It turns out that, when $\theta_{13}$
is choosen to be zero, Eq. (\ref{aeq_2}) and the first one of Eqs. (\ref{m_eqs})
are protected by the requirement that $\theta_{23} = 45^o$.
Breaking relation (\ref{aeq_1}) will give a mass
to the second neutrino state, which can account for the splitting
necessary for solar neutrino oscillations.
An interesting note is that if the second relation in (\ref{m_eqs}) remains
unchanged, then the solar mixing angle will be given by
$\tan^2 \theta_{12} = 0.5$, which is very close to the best fit value for
solar neutrino oscillations.

With these choices, the following neutrino Dirac mass matrix is obtained:
$$
m_D = \left(
\begin{array}{ccc}
m_1 & m_1 + \frac{\delta}{\sqrt{2}}  &
 m_1 + \frac{\delta}{\sqrt{2}} \\
  & m_3 & 2 m_1 -m_3 \\
 &  & m_3
 \end{array}
\right)
$$ (the matrix being symmetric). This texture gives rise
 the following neutrino
masses:
$$ \{ m_{\nu_0}, m_{\nu_1}, m_{\nu_2}, m_{\nu_3} \}
= \{ 3 \frac{c^2 + 3 m_1^2}{m_M}\ , \
 0 \ , \  \delta +  {\cal O}(\delta \frac{m_1^2}{c^2}) \ ,\
  4 \frac{(m_1 - m_3)^2}{M_M} \}
$$
and the 4-neutrino mixing matrix will be:
$$
U_{\alpha i} = \left(
\begin{array}{cccc}
\frac{c^2}{c^2 + 3 m_1^2} & 0 & 0 & 0 \\
\frac{m_1\ +\ \sqrt{2}/3\ \sqrt{\delta m_M} }{c} & \sqrt{\frac{2}{3}} & \sqrt{\frac{1}{3}} & 0 \\
\frac{m_1\ +\ \sqrt{2}/6\  \sqrt{\delta m_M} }{c} & \sqrt{\frac{1}{6}} & -\sqrt{\frac{1}{3}} & -\frac{1}{\sqrt{2}} \\
\frac{m_1\ +\ \sqrt{2}/6\  \sqrt{\delta m_M} }{c} & \sqrt{\frac{1}{6}} & -\sqrt{\frac{1}{3}} & \frac{1}{\sqrt{2}}
 \end{array}
\right)
$$
(with corrections of order $m_1/c$). Here $m_1$ can be taken to be
arbitrary (as long as it is smaller than $c$), and $m_3$ to be
fixed by the mass of the third neutrino: $ m_3  = \pm \sqrt{M_M
m_{\nu_3}}/2 + m_1$.

 We have obtained the hierarchical structure for
neutrino masses. Taking $m_{\nu_3} = M \simeq 5. \times 10^{-2}$ eV and
$m_{\nu_2} = \delta
\simeq 7. \times 10^{-3}$ eV, the splitting between $\nu_3$ and
$\nu_2$ accounts for atmospheric neutrino oscillations, while the
splitting between $\nu_2$ and $\nu_1$ accounts for the solar
neutrinos. The atmospheric mixing angle is $45^o$, the solar angle is
large ($\simeq  35^o$), and the $\theta_{13}$ angle is zero. Choosing $c
= 1$ TeV for $m_M= 10^{15}$ GeV, and $m_1 = 0.11 c$, we obtain
the mass of the fourth neutrino $m_{\nu_0} \simeq 3 $eV, and the
elements of the rotation matrix:
$$
U_{e 0} \simeq 0.14 \ ,\ U_{\mu 0} = U_{\tau 0} \simeq 0.12 \ ,
$$
which values can account for the LSND results.

\subsection {Massless neutralino  and  large mixing}

Another interesting question is if it is possible to explain the large mixing
among the three SM flavor eigenstates through the coupling to the neutralino.
To this purpose, let's assume that the neutrino Dirac mass matrix has
a diagonal structure:
$$ (m_D)_{\alpha \beta} = m_{\alpha} \delta_{\alpha \beta} $$
Let's, moreover, assume that $M_{\nu}(\chi^0,\chi^0)  = (c_1^2 + c_2^2 + c_3^2)/m_M$
is much  larger than the neutrino masses
$m_{\alpha}^2/m_M$. Then, we can decouple
the fourth neutrino, and in the see-saw approximation the mass matrix for
the three SM neutrinos will become:
$$ (M_{\nu})'_{\alpha \beta} \simeq \frac{ m_{\alpha} m_{\beta}}{m_M} \left(
\delta_{\alpha \beta} - \frac{c_{\alpha} c_{\beta}}{\vec{c}^2}
\right)
$$
with $\vec{c}^2 = c_1^2 + c_2^2 + c_3^2$. Note that this matrix has zero
determinant, but it is not traceless. As a consequence, the only possible
diagonal neutrino mass matrices which can be obtained by diagonalization
are the hierarchical diag$(0,\delta,M)$ structure and the
inverted hierarchy case diag$(M,M+\delta,0)$. Both of these require that
$$ |(M_{\nu}')_{12}| \simeq |(M_{\nu}')_{13}| \simeq 0$$
$$ |(M_{\nu})'_{22}| \simeq |(M_{\nu}')_{33}| \simeq |(M_{\nu}')_{23}| \simeq M/2$$
with corrections to these equations of order $\delta/M$. A simple choice
which satisfies the above equations is
\begin{equation}\label{cm_c}
 (c_1,c_2,c_3) = (0,c,c) \ ; \ (m_1,m_2,m_3) = (0,\sqrt{M m_M},\sqrt{M m_M})
\end{equation}
which leads to a neutrino mass matrix:
\begin{equation}\label{hier}
 M_{\nu}' =
\left(
\begin{array}{ccc}
0 & 0 & 0 \\
0 & M/2 & -M/2\\
0 & -M/2 & M/2
 \end{array}
\right).
\end{equation}
This is the hierarchical case, with $m_{\nu_3} = M$ and
the angles $\theta_{23} = 45^o$, $\theta_{13} = 0$.
Corrections of order $\delta/M$ to Eq. (\ref{cm_c}):
\begin{equation}\label{cm_c2}
 (c_1,c_2,c_3) = c( \sqrt{\frac{4 \delta}{3 M}},
 1-\frac{\delta}{3M},1-\frac{\delta}{3M})
 \ ;
\end{equation}
$$
\ (m_1,m_2,m_3) = (\sqrt{\delta m_M/3},\sqrt{M m_M},\sqrt{M m_M})
$$
will provide a mass $\delta$ for $\nu_2$ and make the $\theta_{12}$ angle
$\simeq 35^o$.

Finally, we want to mention that using a sterile neutrino in order
to explain large mixing in the active sector is not a new idea
(see, for example \cite{balaji}). However, while in \cite{balaji}
the couplings of the sterile neutrino with the three SM flavors is
introduced in a somewhat {\it ad hoc} manner, in our model it
arises naturally.

\section{Conclusions}

In this paper we have addressed two questions: is a massless
neutralino allowed by experimental data? and can such a neutralino
couple with the neutrino sector and explain the LSND result,
and/or large mixing among the three active neutrino flavors? We
answered both questions in the affirmative. By giving up the
assumption of universal gaugino masses, we can obtain a massless
neutralino in the MSSM. This would require fine-tuning, but, since
its couplings to the $Z$ can be suppressed, it will still satisfy
the LEP constraints on the invisible $Z$ width. Conversely, we can
obtain a massless neutralino in the framework of the NMSSM, in
which case this particle will be mostly a singlino.

The coupling of the massless neutralino with the neutrino sector is achieved
through the introduction of new soft SUSY breaking terms of the form
$ c \tilde{B} N$. Then the neutralino, which is identified
with a sterile neutrino, will aquire a mass
proportional to the square of the soft SUSY breaking scale
($c^2/M_M \sim $eV), in contrast with the usual see-saw where the light
neutrino mass is proportional to the up quark or charged lepton mass
square. This makes it an ideal candidate for explaining the LSND
experiment results. Moreover, the see-saw induced mixing of the neutralino
with the three neutrinos is also consistent with constraints from short
baseline experiments, while large enough to account for LSND. Large
mixing among the three active neutrino flavors arises naturally in this model.

The weak point of this model is that fine tuning is required to
obtain a massless neutralino. A promising candidate for a massless
SUSY particle would be a fermion associated with spontaneous SUSY
breaking, that is, a visible sector goldstino. However, this would
require some more model building, and we leave it for another
paper. Even so, we find the fact that a massless neutralino is
experimentally allowed interesting. Also, the fact that one of the
light neutrino masses can be connected to the soft SUSY breaking
scale is another intriguing feature of this scenario.

\section{Acknowledgements}

We thank  K.S. Babu, B. Bajc, Z. Berezhiani, J.L. Chkareuli,
P. Kumar and
H. B. Nielsen for useful discussions. I.G., C.M. and S.N.
gratefully acknowledge
support from the Fermilab Summer Visitor Program and warm
hospitality from the Fermilab Theory Group during the initial
stages of this project. S.N. also acknowledges the warm
hospitality and support of the DESY and CERN Theory Groups
during the completion of this work.
I.G., C.M. and S.N.'s work is supported in part by the U.S.
Department of Energy Grant Numbers DE-FG03-98ER41076 and
DE-FG02-01ER45684. The work of J.L. was supported
by the U.S.~Department of Energy Grant DE-AC02-76CHO3000.


\begin{thebibliography}{99}

\bibitem{ref1}
The SuperKamiokande collaboration, Phy. Rev. Lett. {\bf 85}, 3999
(2000) and  Phys. Rev. Lett. {\bf 82}, 2644 (1999);
 T. Toshito, for the SK collaboration, hep-ex/0105023; The
MACRO collaboration, Phy. Lett, B517, 59 (2001); G. Giacomelli and
M. Giorgini, for the MACRO collaboration, hep-ex/0110021.

\bibitem{ref2}
 The {\sc Gallex}
collaboration,  Phy. Lett. {\bf B447}, 127 (1999); The SAGE
collaboration, Phys. Rev.  {\bf C60}, 055801 (1999); The
SuperKamiokande collaboration, hep-ex/0103032 {\rm and}
Phys. Rev Lett. {\bf 86}, 5651 (2001); 
The GNO collaboration, Phy. Lett. {\bf B490}, 16
(2000); The SNO collaboration, Phys. Rev. Lett.  {\bf 87}, 71301
(2001); Q.~R.~Ahmad {\it et al.}  [SNO Collaboration],
nucl-ex/0204008 and nucl-ex/0204009.

\bibitem{sees}
M. Gell-Mann, P. Ramond and R. Slansky, in {\it Supergravity},
Proceedings of the workshop, Stony Brook, New York, 1979, edited
by P. van Nieuwnehuizen and D. Freedman (North-Holland, Amsterdam,
1979), p.315; T. Yanagida, in {\it Proceedings of the Workshop on
the Unified Theories and Baryon Number in Universe}, Tsukuba,
Japan, 1979, edited by O. Sawada and A. Sugamoto (KEK Report No.
79-18), Tsukuba, 1979), p.95; R.N. Mohapatra and G. Senjanovi\'c,
Phys. Rev. Lett. {\bf 44}, 912 (1980).

\bibitem{ref5}
C.H.~Albright, K.S.~Babu and S.M.~Barr, Nucl. Phys. Proc. Suppl
{\bf 77}, 308 (1999); G.~Altarelli and F.~Feruglio, Phys. Lett.
{\bf B439}, 112  (1998). Z. Berezhiani, Z. Tavartkiladze, Phys.
Lett. {\bf B409}, 220 (1997); Z. Berezhiani and A. Rossi, JHEP
{\bf 9903}, 002 (1999); Q. Shafi and  Z. Tavartkiladze, Phys. Lett. {\bf
451}, 129 (1999);  Phys. Lett. {\bf 482}, 145 (2000).


\bibitem{ref6}
K.S.~Babu and S.M.~Barr, Phys. Lett. {\bf B381}, 202 (1996); Z.
Berezhiani, A. Rossi, Phys. Lett. {\bf B367}, 219 (1996); S. F.
King, Nucl. Phys. {\bf B562}, 57 (1999); G. Altarelli, F.
Feruglio, hep-ph/0206077 and references therein.

\bibitem{ref7}
A.Yu.~Smirnov, Phys. Rev. {\bf D48}, 3264 (1993).

\bibitem{ref8}
K.S. Babu, C.N. Leung and J. Pantaleone, Phys. Lett. {\bf B319},
191 (1993); P.H. Chankowski and Z. Pluciennik, Phys. Lett. {\bf
B316}, 312 (1993); J. Ellis and S. Lola, Phys. Lett. {\bf B458},
310, (1999); K.R.S. Balaji, A.S. Dighe, R.N. Mohapatra and M.K.
Parida, Phys. Rev. Lett. {\bf 84}, 5034 (2000); Phys. Lett. {\bf
B481}, 33 (2000); P.H.~Chankowski, A.~Ioannisian, S.~Pokorski,
J.W.F~Valle, Phys.Rev.Lett. {\bf 86}, 3488 (2001)  and references
therein.


\bibitem{LSND}LSND Collab., C. Athanassopoulos et.al.,
Phys. Rev. Lett. {\bf 77}, 3082 (1996);
 LSND Collab., A. Aguilar et. al.,
Phys. Rev. {\bf D64}, 112007 (2001).

\bibitem{sss1}
D.O. Caldwell and R.N. Mohapatra, Phys. Rev. {\bf D48}, 3259
(1993); J. Peltoniemi, J.W.F. Valle, Nucl. Phys. {\bf B406}, 409
(1993); 
V.D. Barger, S. Pakvasa, T.J. Weiler and K. Whisnant, Phys. Rev. {\bf D58},
093016 (1998);
 S.C. Gibbons, R.N. Mohapatra, S. Nandi and A. Raychoudhuri, Phys.
Lett. {\bf B430}, 296 (1998);
 K.S. Babu and  R.N. Mohapatra,
Phys.Lett. {\bf B522} 287 (2001); B.Brahmachari, S. Choubey , R.
N. Mohapatra, Phys.Lett. {\bf B536}, 94 (2002).

\bibitem{balaji}
K.R.S. Balaji, A. Perez-Lorenzana, A. Yu. Smirnov,
Phys.Lett. {\bf B509},  111 (2001).


\bibitem{cpt}
H.~Murayama and T.~Yanagida, Phys. Lett.  {\bf B520}, 263 (2001);
G.~Barenboim, L.~Borissov, J.~Lykken and A.~Y.~Smirnov, JHEP
{\bf 0210}, 001 (2002); G. Barenboim, L. Borissov , J. Lykken, Phys. Lett.
{\bf B534}, 106 (2002); G. Barenboim, J.F. Beacom, L. Borissov,
B. Kayser,  Phys. Lett. {\bf B537}, 227 (2002); A.~Strumia,
hep-ph/0201134; A.~de ~Gouvea, hep-ph/0204077; 
G.~Barenboim and J.~Lykken, hep-ph/0210411.



\bibitem{grish} L. Girardello and M. T. Grisaru, Nucl.Phys. {\bf B194}, 65
(1982); I. Jack and D.R.T. Jones, Phys.Lett. {\bf B457}, 101
(1999); S. P. Martin, Phys.Rev. {\bf D61}, 035004 (2000).

\bibitem{hk}H. E. Haber and G. L. Kane, Phys.Rept.{\bf 117}, 75
(1985); H.P. Nilles, Phys. Rep.{\bf 110}, 1 (1984).

\bibitem{ster}
A. Barl, H. Fraas, W. Majerotto and N. Oshimo, Phys. Rev. {\bf D
40}, 1594 (1989),  B.R. Kim, A. Stephan and S.K. Oh, Phys. Lett.
{\bf B336}, 200 (1994); F. Franke, H. Fraas and A. Bartl,
Phys. Lett. {\bf B336}, 415 (1994); F. Franke and H. Fraas, Int.J.
Mod. Phys. {\bf A12}, 479 (1997).

\bibitem{pdg}Particle Data Group, Phys. Rev. {\bf D66}, 01001 (2002).


\bibitem{bugey} Bugey Collab., B. Achkar et. al.,
Nucl. Phys. {\bf B434}, 503 (1995).

\bibitem{CHDS} CHDS Collab., F. Dydak et. al.,
Phys.Lett. {\bf B134}, 281 (1984).

\bibitem{CHOOZ} CHOOZ Collab., M. Appolonio et. al.,
Phys. Lett. {\bf B420}, 397 (1998).

\bibitem{Babu_bm} K.S. Babu and R.N. Mohapatra,
Phys. Lett. {\bf B532}, 77 (2002).


\end{thebibliography}
\end{document}